\documentclass[twoside,twocolumn,9pt]{article}
\usepackage{extsizes}
\usepackage[super,sort&compress,comma]{natbib} 
\usepackage[version=3]{mhchem}
\usepackage[left=1.5cm, right=1.5cm, top=1.785cm, bottom=2.0cm]{geometry}
\usepackage{balance}
\usepackage{packages/widetext}
\usepackage{times,mathptmx}
\usepackage{sectsty}
\usepackage{graphicx} 
\usepackage{lastpage}
\usepackage[format=plain,justification=raggedright,singlelinecheck=false,font={stretch=1.125,small,sf},labelfont=bf,labelsep=space]{caption}
\usepackage{float}
\usepackage{fancyhdr}
\usepackage{fnpos}
\usepackage[english]{babel}
\usepackage{array}
\usepackage{droidsans}
\usepackage{charter}
\usepackage[T1]{fontenc}
\usepackage[usenames,dvipsnames]{xcolor}
\usepackage{setspace}
\usepackage[compact]{titlesec}
\usepackage{subfigure}
\usepackage[normalem]{ulem}

\usepackage{epstopdf}

\definecolor{cream}{RGB}{222,217,201}

\begin{document}

\pagestyle{fancy}
\thispagestyle{plain}
%

\makeatletter
\renewcommand\LARGE{\@setfontsize\LARGE{15pt}{17}}
\renewcommand\Large{\@setfontsize\Large{12pt}{14}}
\renewcommand\large{\@setfontsize\large{10pt}{12}}
\renewcommand\footnotesize{\@setfontsize\footnotesize{7pt}{10}}
\makeatother

\renewcommand{\thefootnote}{\fnsymbol{footnote}}
\renewcommand\footnoterule{\vspace*{1pt}%
\color{cream}\hrule width 3.5in height 0.4pt \color{black}\vspace*{5pt}} 
\setcounter{secnumdepth}{5}

\makeatletter 
\renewcommand\@biblabel[1]{#1}            
\renewcommand\@makefntext[1]%
{\noindent\makebox[0pt][r]{\@thefnmark\,}#1}
\makeatother 
\renewcommand{\figurename}{\small{Fig.}~}
\sectionfont{\sffamily\Large}
\subsectionfont{\normalsize}
\subsubsectionfont{\bf}
\setstretch{1.125} 
\setlength{\skip\footins}{0.8cm}
\setlength{\footnotesep}{0.25cm}
\setlength{\jot}{10pt}
\titlespacing*{\section}{0pt}{4pt}{4pt}
\titlespacing*{\subsection}{0pt}{15pt}{1pt}

\fancyfoot{}
\fancyhead{}
\renewcommand{\headrulewidth}{0pt} 
\renewcommand{\footrulewidth}{0pt}
\setlength{\arrayrulewidth}{1pt}
\setlength{\columnsep}{6.5mm}
\setlength\bibsep{1pt}

\makeatletter 
\newlength{\figrulesep} 
\setlength{\figrulesep}{0.5\textfloatsep} 

\newcommand{\topfigrule}{\vspace*{-1pt}%
\noindent{\color{cream}\rule[-\figrulesep]{\columnwidth}{1.5pt}} }

\newcommand{\botfigrule}{\vspace*{-2pt}%
\noindent{\color{cream}\rule[\figrulesep]{\columnwidth}{1.5pt}} }

\newcommand{\dblfigrule}{\vspace*{-1pt}%
\noindent{\color{cream}\rule[-\figrulesep]{\textwidth}{1.5pt}} }

\makeatother

\twocolumn[
  \begin{@twocolumnfalse}
\vspace{4cm}
\sffamily
\begin{tabular}{m{4.5cm} p{13.5cm} }

& \noindent\LARGE{\textbf{Cell division: a source of active stress in cellular monolayers$^\dag$}} \\
\vspace{0.3cm} & \vspace{0.3cm} \\

 & \noindent\large{Amin Doostmohammadi,\textit{$^{a}$} Sumesh P. Thampi,\textit{$^{a}$} Thuan B. Saw,\textit{$^{b}$} Chwee T. Lim,\textit{$^{b,c}$} Benoit Ladoux,\textit{$^{b,d}$} and Julia M. Yeomans$^{\ast}$\textit{$^{a}$}} \\

& \noindent\normalsize{We introduce the notion of cell division-induced activity and show that the cell division generates extensile forces and drives dynamical patterns in cell assemblies. Extending the hydrodynamic models of lyotropic active nematics we describe turbulent-like velocity fields that are generated by the cell division in a confluent monolayer of cells. We show that the experimentally measured flow field of dividing Madin-Darby Canine Kidney (MDCK) cells is reproduced by our modeling approach. Division-induced activity acts together with intrinsic activity of the cells in extensile and contractile cell assemblies to change the flow and director patterns and the density of topological defects. Finally we model the evolution of the boundary of a cellular colony and compare the fingering instabilities induced by cell division to experimental observations on the expansion of MDCK cell cultures.} \\

\end{tabular}

 \end{@twocolumnfalse} \vspace{0.6cm}

  ]

\renewcommand*\rmdefault{bch}\normalfont\upshape
\rmfamily
\section*{}
\vspace{-1cm}


\footnotetext{\textit{$^{a}$~The Rudolf Peierls Centre for Theoretical Physics, 1 Keble Road, Oxford, OX1 3NP, UK. Tel: +44 01865 273952; E-mail: julia.yeomans@physics.ox.ac.uk}}
\footnotetext{\textit{$^{b}$~Mechanobiology Institute, National University of Singapore, Singapore 117411, Singapore. }}
\footnotetext{\textit{$^{c}$~Department of Biomedical Engineering, National University of Singapore, Singapore 117575, Singapore. }}
\footnotetext{\textit{$^{d}$~Institut Jacques Monod, CNRS UMR 7592, Université Paris Diderot, Paris 75013, France. }}

\footnotetext{\dag~Electronic Supplementary Information (ESI) available: [details of any supplementary information available should be included here]. See DOI: 10.1039/b000000x/}



\section{Introduction}
The collective migration of cells plays a pivotal function in vital physiological processes such as embryonic morphogenesis \cite{Haas2006}, tissue development \cite{Ghosh2007}, wound healing \cite{Silberzan2007,Wolgemuth2011}, and tumor growth \cite{Wolf2007,Cheng2009}. The collective motion of cells is often explained by considering a delicate interplay between biochemical signaling, metabolic processes, and mechanical forces \cite{Tse2012}. However, the individual function played by each of these mechanisms and their relative importance remain obscure. 

The sources of mechanical stimuli often originate within the cell culture. Examples are cell motility due to molecular motors deforming the cytoskeleton, and flow fields established by cell division. In the past decade, there has been growing evidence of the important impact of mechanical processes such as motility-induced forces in the growth and development of cellular cultures \cite{Lo2000,Tan2003,Trepat2009,Vedula2014}. The collective migration of cancer cells due to the generation of compressive stress during tumor growth \cite{Tse2012}, the coordination of cell motion by the emergence of a mechanical wave during epithelial monolayer expansion \cite{Trepat2012}, and the emergence of fingering instabilities at the border of a growing tissue \cite{Silberzan2007}, are striking examples. Nevertheless, it is still unclear how the collective motion is induced by mechanical stresses generated at the scales of individual cells. 

Recently, important correlations have been found between mechanical forces within cell cultures and cell proliferation \cite{Basan2013,Campinho2013}. It has been shown that the mechanical forces generated by cell division and apoptosis can result in fluidization of tissue \cite{Joanny2010} and lead to the formation of long-range vortical structures in living tissue \cite{Rossen2014}. The dependence of cell division rates on internal mechanical stress has been suggested as a possible mechanism for growth control \cite{Shraiman2005} and cellular flow generated by proliferation gradients has been shown to be drastically suppressed by mechanical stresses \cite{Joanny2013}. Continuum theories have been used to model anisotropic tissue growth with active stresses generated by cell division \cite{Julicher2008,Julicher2009}. Despite the widespread implications, the contribution of cell division to coordinated cell motion remains largely unexplored.  

In this paper, we investigate the generation of active stresses by cell division. Using generalised equations of nematic liquid crystals with the concentration of nematics allowed to vary, we reproduce the experimental flow field of dividing Madin-Darby Canine Kidney (MDCK) cells and show that a cell division leads to a dipole-like flow field at the division site. Even in the absence of intrinsic activity of the cells, such division-induced activity can generate turbulent-like flows. We study the role of cell division on the cellular flow and compare the effects of division-induced activity in assemblies of extensile and contractile cells. The role of cell division in determining the growth of an interface bounding a layer of cells is also investigated. We find fingering instabilities at the cellular front, which are compared to experiments on a growing colony of MDCK cells.
\section{Methodology}
\subsection{Equations of motion}
There are many models for collective cell migration in the literature \cite{Gruler1999,Byrne2009,Fozard2010,Landman2013,Hakim2013}. The cultures of cells are often regarded as a continuous medium due to strong contacts between the cells \cite{Joanny2010,Rossen2014}. Here, in order to describe the dynamics of the collective motion of cells, we employ a continuum approach based on the hydrodynamic equations of active nematics \cite{Davide2007,Marchetti2013}
\begin{align}
(\partial_t + u_k \partial_k) Q_{ij} - S_{ij} = \Gamma H_{ij},
\label{eqn:lc}\\
\partial_{t}\rho+\partial_{i}(\rho u_{i})=0,
\label{eqn:cont}\\
\rho (\partial_t + u_k \partial_k) u_i = \partial_j \Pi_{ij},
\label{eqn:ns}
\end{align}
where $Q_{ij}=2q(n_{i}n_{j}-\delta_{ij}/2)$ is the two-dimensional nematic order tensor, with  director ${\bf n}$ and magnitude $q$, taken to characterise the coarse-grained orientation of cells, which is a consequence of the underlying cytoskeletal network. Based on the comparison of the flow field generated in our simulations with that of the experiments, the length scale of a cell corresponds to approximately six grid points in the continuum formulation.

Equation~(\ref{eqn:lc}) is a convection-diffusion equation for the dynamics of the order parameter field $\mathbf{Q}$. The total density $\rho$ obeys the continuity equation~(\ref{eqn:cont}).
The evolution of the velocity field {\bf u} is described by equation~(\ref{eqn:ns}), which is the generalisation of the Navier-Stokes equations to describe the dynamics of liquid crystals.

Consider first the evolution of the order parameter field,  equation~(\ref{eqn:lc}). As a result of the anisotropic shape of the nematogens, the director field responds to the gradients in the flow field. This is accounted for by the generalised advection term
\begin{align}
 S_{ij} =& (\lambda E_{ik} + \omega_{ik})(Q_{kj} + \delta_{kj}/3)+ (Q_{ik} + \delta_{ik}/3)(\lambda E_{kj} - \omega_{kj})\nonumber\\
 &- 2 \lambda (Q_{ij} + \delta_{ij}/3)(Q_{kl}\partial_k u_l),
 \label{eqn:corr}
\end{align}
where
\begin{equation}
E_{ij} = (\partial_i u_j + \partial_j u_i)/2, \quad \quad \quad \quad \omega_{ij} = (\partial_j u_i - \partial_i u_j)/2
\end{equation}
are the symmetric and antisymmetric parts of the velocity gradient tensor, respectively called the strain rate and vorticity tensors.

In equation~(\ref{eqn:corr}), $\lambda$ is the alignment parameter which determines whether the collective response of the nematogens to a velocity gradient is dominated by the strain or vorticity. The director aligns at a given angle to a shear flow, called the Leslie angle, if
\begin{equation}
\mid \lambda \mid\;\; > \frac{9q}{3q+4}.
\end{equation}
If this equality is not satisfied the director field rotates (tumbles) in a simple shear flow. Generally, $\lambda$ depends on the shape of the particles, with $\lambda>0$ and $\lambda<0$ corresponding to rod-like and plate-like particles respectively.

The molecular field
\begin{equation}
H_{ij} = -\frac{\delta \mathcal{F}}{\delta Q_{ij}} + \left(\frac{\delta_{ij}}{3}\right) {\rm Tr} \left(\frac{\delta \mathcal{F}}{\delta Q_{kl}}\right)\label{eqn:molpot}\\
\end{equation}
ensures that the system relaxes diffusively to the minimum of a free energy, $\mathcal{F}$, through equation~(\ref{eqn:lc}) and the diffusion constant $\Gamma$ controls the time scale over which the relaxation occurs. 
We use the Landau-de Gennes free energy
\begin{equation}
\mathcal{F} =  \frac{1}{2}A(q^{2}c-\frac{1}{2}Q_{ij}Q_{ji})^{2} + \frac{K}{2} (\partial_k Q_{ij} )^2 
\label{eq:free}
\end{equation}
where $A$ is a material parameter characterising the coupling between the nematic order, ${\bf Q}$, and the concentration of cells, $c$ and $K$ is an elastic constant measuring the energy associated with deviations from nematic ordering. In two dimensions there are, in general, two elastic constants associated with bend and splay deformations, and we take these to be equal. The free energy~(\ref{eq:free}) corresponds to a molecular field
\begin{align}
H_{ij}=-AQ_{ij}\left(q^{2}c-\left(Q_{ij}Q_{ji}\right)/2\right)+K(\partial_{k}^{2}Q_{ij}).
\end{align}
\begin{figure*}[t]
\includegraphics[trim = 0 0 0 0, clip, width = 1.0\linewidth]{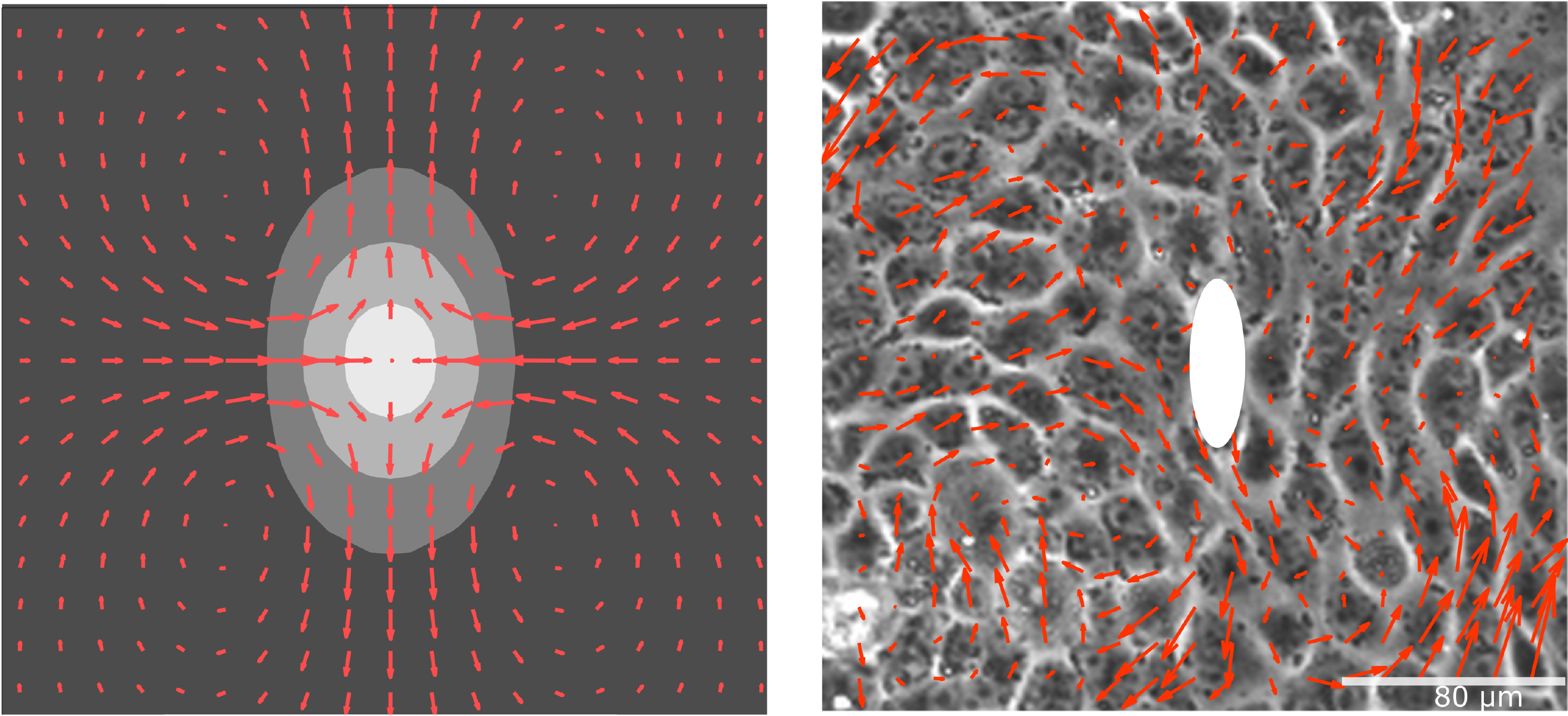}
\caption{Generation of a dipole-like flow field due to a cell division event modelled as a local increase in concentration (left) and experimental measurements of the flow field around a dividing cell (right). Red arrows denote velocity fields and the colourmap in the simulation (left) shows concentration. The position of the dividing cell in the experiment is marked by the white ellipse.}
\label{fig:single}
\end{figure*}
Turning now to the Navier-Stokes equation~(\ref{eqn:ns}), the stress term, $\Pi_{ij}$, includes contributions from the viscous stress
\begin{align}
\Pi_{ij}^{viscous}=2\eta E_{ij},
\end{align}
with the viscosity $\eta$, and elastic stresses
\begin{align}
\Pi_{ij}^{elastic}&=-P\delta_{ij} + Q_{ik}H_{kj} - H_{ik} Q_{kj} - K\partial_i Q_{kl}\partial_{j}Q_{kl}\nonumber\\
&\ \ \ + \lambda\big[2(Q_{ij} + \frac{\delta_{ij}}{3}) (Q_{kl} H_{lk})
- H_{ik} (Q_{kj} + \frac{\delta_{kj}}{3})\nonumber\\
&\ \ \ \ \ \ \ \ \ - (Q_{ik} + \frac{\delta_{ik}}{3}) H_{kj}\big],
\label{eqn:elstress}
\end{align}
where $P$ is the pressure. The elastic stress is a second consequence of the anisotropic nature of the nematogens. It occurs because, as the directors turn they induce stresses and hence a contribution to the velocity field, often called the `back-flow'. In addition, the stress term includes the active stress
\begin{align}
\Pi_{ij}^{active}=-\zeta Q_{ij},
\label{eq:act}
\end{align}
where the coefficient $\zeta$ measures the strength of the activity. In general, the active stress includes contributions from stresses induced by molecular motors and actin polymerisation dynamics. $\zeta<0$ describes a contractile system where the flow field of active nematogens is along their short axis and $\zeta>0$ describes an extensile system where the flow is along their long axis. The form of the active stress tensor is motivated in \cite{Sriram2002}. Here, in order to distinguish the division-induced activity from active stresses in equation (\ref{eq:act}), we refer to the latter as `intrinsic' activity of the cells.

Many properties of active nematics can be interpreted by noting that the active term appears in the stress, under a derivative, and therefore any spatial gradient in the direction or orientation of the nematic order induces a flow. An immediate consequence is that the active nematic state is hydrodynamically unstable\cite{Sriram2002,Joanny2005} and the global nematic ordering is replaced by active turbulence, a state characterised by swirls and jets in the velocity field and topological defects in the director field \cite{Julia2012}.
Equations.~(\ref{eqn:lc})--(\ref{eqn:ns}) have proven successful in modelling the behaviour of dense suspensions of microswimmers and active suspensions of microtubules driven by molecular motors \cite{Giomi2013,ourprl2013}.

To introduce cell division, the equation of motion for the concentration of cells is taken as
\begin{align}
\partial_{t}c+\partial_{i}(u_{i}c)=\kappa\nabla^{2}c+\alpha c,
\label{eqn:conc}
\end{align}
where $\kappa$ is the diffusivity of the cells and $\alpha$ represents the proliferation rate due to the growth of cells. We will show that any increase in concentration results in the generation of active stresses, which drive the flow of cells. Note that the form of the free energy in equation~(\ref{eq:free}) corresponds to a constant concentration throughout the system in equilibrium. In section 5 of the paper we will generalise the free energy and the concentration equation of motion to allow us to model the mechanical behaviour of a dividing cellular colony with a free surface.

The equations of motion~(\ref{eqn:lc})--(\ref{eqn:ns}) and  (\ref{eqn:conc})   are solved using a hybrid lattice Boltzmann method \cite{MatthewPRL}. Simulations were performed in a two-dimensional square domain of size $200\times200$ and discrete space and time steps were chosen as unity. Unless otherwise stated, the parameters used in the simulations are $\rho=40$, $\Gamma=0.1$, $\kappa=0.1$, $K=0.05$, $\alpha=0.0001$, $\lambda=0.3$, $q=1$, $\zeta=0.0025$ and $\eta=2/3$ in Lattice-Boltzmann units \cite{ourprl2013,MatthewPRL}. We prescribe zero velocity field, a uniform concentration and random nematogen orientations as initial conditions and implement periodic conditions on domain boundaries.
\subsection{Experiments}
Madin-Darby canine kidney (MDCK WT) strain II cells were cultured in low glucose DMEM medium (Invitrogen), with $100~\mu g/ml$ penicillin, $100~\mu g/ml$ streptomycin (Invitrogen) and $10\%$ FBS (Invitrogen). Experiments for the cell division flow field measurement (DFF) were performed with cells incubated in Leibovitz's $L-15$ (Invitrogen), with $100~\mu g/ml$ penicillin, $100~\mu g/ml$ streptomycin (Invitrogen) and $10\%$ FBS (Invitrogen). Under these conditions, the typical aspect ratio of cells (if fitted with an ellipse) averaged at $\sim1.8$, and could reach up to $3-4$. $10~\mu M$ blebbistatin (Cayman Chemical Company) drug was added and left in the medium throughout the experiment to slow down dynamics of tissue and reduce noise in measurements. Measurements were done without drug as well, and similar cell-induced vortices were also measured, but happening earlier after cell division event. The fingering experiment (FE) used a culture medium without the drug.

For DFF, MDCK cells were confined on a $500~\mu m$ diameter square pattern by a microcontact printing technique \cite{Vedula2013microfabricated}. Confinement is essential to get a confluent tissue without too much spatial variations in cell density, and also to establish nematic ordering of cells, since tissue will be polarized if left to expand into free area. Fibronectin (FN - $25~\mu g/ml$ Atto dye conjugated FN and $100~\mu g/ml$ pure FN, Sigma and Roche) was incubated on polydimethylsiloxane PDMS (Sylgard $184$; Dow Corning) stamp for $1~hr$ at room temperature, dried and stamped on a PDMS spin-coated dish. Pluronics F127 (Sigma) $1\%$ was incubated to passivate the non-stamped area, before cells were seeded and left overnight to grow to confluence on the stamped area and imaging the next day. 
For FE, cells were seeded overnight on a FN (pure FN, $20~\mu g/ml$, $1~hr$ incubation) incubated glass bottom dish (Ibidi). A PDMS block with long, straight boundaries was placed on the surface before cell seeding to prevent cells from penetrating the desired area.  After cells outside the block grew to confluence, the block was lifted and imaging started when cells started to migrate into the void, forming fingers. 

Time-lapse Imaging was done respectively for $\sim 2~days$ (DFF) and $\sim 10~hrs$ (FE) (Biostation (Nikon)). Imaging resolution was $0.647~\mu m/pixel$ with a $10X$ Phase objective. The image interval was $10~min$.

The velocity field of a confluent tissue in the confined region was measured \cite{Silberzan2010,Doxzen2013}. Individual cell division events were pinpointed by eye. The cell division starting point was chosen to be the last frame where the mother cell was still present. Each division starting point was chosen to be a new reference point ($0,0$) and the velocity around this point was determined and rotated with respect to the closest nematic director, such that the nematic director aligned in the $y$-axis in this new reference frame. The transformed velocity field of division event, $i$, $f$ time-frames after the division starting point is:  ${\bf v}_{i,f}$ (the origin of the local reference frame at time-frame, $f$, was always chosen to be the same position as the division starting point i.e. at $f=0$). The average transformed velocity field was determined for different $f\in{0,1,2, ...}$, as ${\bf V}_{f}(r)=\sum_{i}^{N}{\bf v}_{i,f}/N$. The drift corrected, average transformed velocity field is ${\bf V}_{f,ND}(r)={\bf V}_{f}(r)-\sum_{r}{\bf V}_{f}(r)/N_r$, where $N_r$ is number of local velocity vectors around each new origin of area $120\times120~\mu m^{2}$ (one cell length is $\sim 15~\mu m$).
\section{Division-induced flow field and meso-scale turbulence}

The flow field generated by the coordinated motion of cells exhibits turbulent-like flow patterns characterised by a distribution of flow vortices, and often referred to as meso-scale or active turbulence \cite{Silberzan2010,Julia2012,Jorn2013}. The emergence of meso-scale turbulence in cellular matter is induced by the motion of the cells which is driven by rearrangements of the cytoskeletal elements by molecular motors. This is modeled by the active term, $\Pi_{ij}=-\zeta Q_{ij}$, in the equations of motion. In order to separate the effects of the cell division from the intrinsic activity of the cells, we first consider the dynamics of a concentration of cells without any conventional active forcing ($\zeta=0$), but proliferating due to cell division.

We present results for the flow field around a single dividing cell, and then describe the velocity field produced by many dividing cells, at random positions and times, but with a given number of divisions, $\phi_{d}$ per unit area at any time.
Cell division is modelled by locally increasing the concentration (by $\alpha$) within a circular area of radius three grid points. The increase in concentration is maintained for a short duration of ten lattice Boltzmann time steps, much smaller than typical evolution time scales of flows in the simulations ($\sim 10,000$). Division events are introduced at random positions as a transient, small, local increase in concentration. This locally changes the nematic ordering, because of the coupling in the free energy, which in turn drives the flow field. We take measurements once the flow has reached a statistically steady state.
\subsection{Flow field of a division event}
Recent experimental observations of the flow fields of endothelial cells have reported the emergence of velocity vortices around cell division sites \cite{Rossen2014}. In these experiments the flow field generated by the cell division may be affected  by a number of mechanical factors such as the motion of the cells and strong friction with the substrate. It has been previously argued that cell division can be thought as a local source of energy injection \cite{Rossen2014} and stress generation \cite{Volfson2008} and that can be modeled as a force dipole \cite{Joanny2010}.  The force dipole can be generated as an elongated cell exerts anisotropic forces on its neighboring cells. Here we show that the flow field is automatically generated in our formulation by a local increase in the concentration, which induces a local active stress. This occurs because, as the concentration is increased, the free energy drives a corresponding increase in the local nematic order (towards $q_{nem}=\sqrt{cq^2}$) with a time lag of the order $t_{q}\sim 1/\Gamma$. Though the concentration disturbance is isotropic, the nematic symmetry of the director field breaks the isotropy. Changes in the local nematic order lead to the variations in the molecular field, $H_{ij}$, and analysis of eqn~(\ref{eqn:elstress}) shows that terms proportional to $\lambda Q_{ij}$ dominate in producing an anisotropic elastic stress.

The flow field generated by a single cell division event is compared with the experimentally measured flow field around a division point in MDCK cells (Fig.~\ref{fig:single}). The flow field reported in Fig.~\ref{fig:single} is after $200$ time steps. In experiments, velocity fields were measured after $180~min$ from the cell division and are averaged over $100$ cell division events. The flow field resembles that of a force dipole with an octupole correction, which takes into account the finite extent of the local increase in the concentration. Moreover, as evident from the simulation results in Fig.~\ref{fig:single}, the division redistributes the concentration along the director orientation and results in the elongation of the cell parallel to the division axis. This is consistent with a recent experiment on MDCK cells, which showed that by orientating along the long axis of the cell, division redistributes the mass along the long axis and causes the outward (inward) movement of neighboring junctions parallel (perpendicular) to the division axis \cite{Wyatt2015}. We note that the spatial length scale of the flow field generated by a single division event in MDCK cells (with reduced motility) spans $\sim 100~\mu m$, similar to the velocity correlation lengths measured in MDCK tissue during normal proliferation and motility \cite{Vedula2012}. Taken together, our simple modeling approach captures important features of the division event and shows that each division acts as a source of activity, generating hydrodynamic flows and driving the system out of thermodynamic equilibrium.
\subsection{Meso-scale turbulence of dividing cells}

\begin{figure*}[tdp]
\begin{centering}
{\includegraphics[trim = 0 0 0 0, clip, width = 0.7\linewidth]{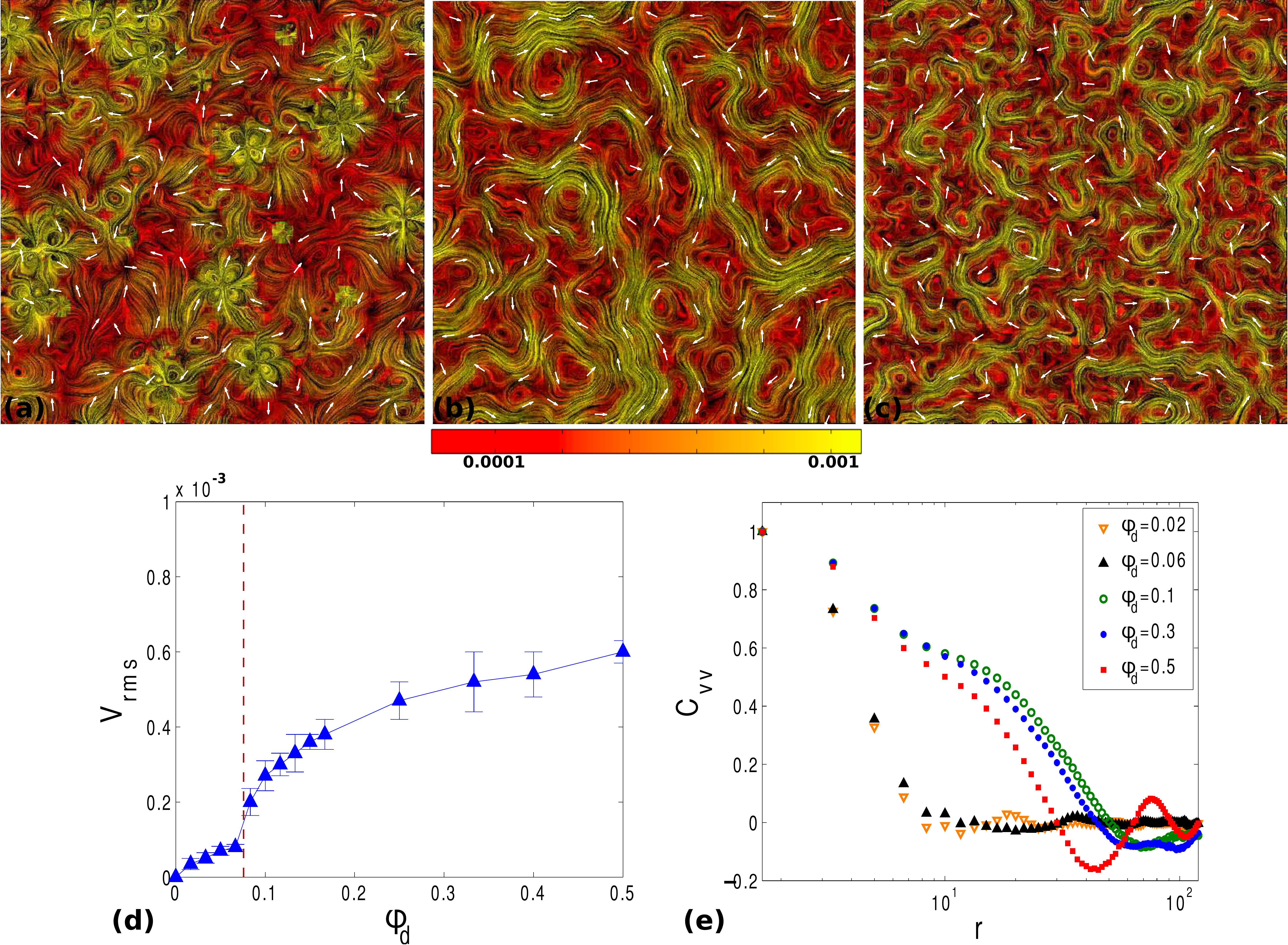}}
\caption{Steady state flow fields for a concentration of dividing cells with no intrinsic activity ($\zeta=0$). (a), (b) and (c) correspond to low $\phi_{d}=0.05$, moderate $\phi_{d}=0.3$, and high $\phi_{d}=0.5$ division fractions. The flow fields are visualised as streamlines using the Line Integral Convolution (LIC) method, the colour maps indicate the magnitude of the velocity and white arrows the local direction of the velocity. We use the same resolution as the flow field simulations to construct LIC maps. (c) The rms-velocity of cell assemblies increases with increasing the division fraction $\phi_{d}$. The sharp increase in $v_{\mbox{rms}}$ at $\phi_{d}= 0.08$ corresponds to the onset of collective behaviour. Panel (d) shows normalised velocity-velocity correlation functions for different division fractions.}
\label{fig:div}
\end{centering}
\end{figure*}
Figure~\ref{fig:div} illustrates the flow field of an assembly of dividing cells in the absence of any intrinsic activity ($\zeta=0$). The activity induced by division events drives the system out of equilibrium and establishes a spatially and temporally evolving flow field. At a low division fraction the flow fields generated by the dividing cells are independent, they do not interact each other. (Fig.~\ref{fig:div}a)).
At division fraction of $\phi_{d}=0.3$, however, the division-induced activity results in the formation of persistent regions of high velocities in straight and curved patterns. The emergence of these jets and vortices in the flow field shows that the cell division triggers coordinated patterns of cellular motion (Fig.~\ref{fig:div}(b)). The dynamics is reminiscent of `active turbulence', the emergence of collective motion in wet active nematics driven by nonzero intrinsic activity ($\zeta\ne 0$) \cite{Dogic2012}. As the division fraction is further increased to $\phi_{d}=0.5$ the swirls and jets are still seen, but the length scale of the vortices decreases (Fig.~\ref{fig:div}(c)).

To quantify these results we measured the root-mean-squared (rms) velocity, and the velocity-velocity correlations function as a function of the cell division fraction. Results for the rms velocity are presented in 
Fig.~\ref{fig:div}(d). As expected, the rms velocity increases as the number of cell divisions increases. Note, however, that there is a sharp increase in rms velocity at $\phi_{d} \equiv \phi^{*}_{d}=0.08$ (red vertical line in Fig.~\ref{fig:div}(d)). This is the cell division fraction beyond which collective patterns and mesoscale turbulence are observed.

The transition to mesoscale turbulence is also evident from measurements of velocity-velocity correlation function, $C_{vv}=\langle\left({\bf v}(r,t).{\bf v}(0,t)\right)/{\bf v}(0,t)^{2}\rangle$, where $\langle\rangle$ denotes temporal and spatial averaging (Fig.~\ref{fig:div}(e)). Below the critical division fraction $\phi_{d}^{\ast}$ the correlation length corresponds to that of  a single division event. Above $\phi_{d}^{\ast}$, however, the correlation length markedly increases corresponding to the emergence of collective behaviour. Further increases in $\phi_{d}$ result in a gradual reduction of the correlation length as the effective activity of the cell assembly increases and the jets and swirls shrink in size. 

Although, several mechanical processes such as a bimodal distribution of cell behavior \cite{Omelchenko2003}, compressive pressure \cite{Butcher2009}, and plithotaxis (cells actively generating heterogeneous intercellular forces and migrating in the direction of maximal principal stress \cite{Trepat2009,TrepatRev}), have been introduced as potential mechanisms of the collective behaviour of cells, the underpinning mechanical source of this behaviour is not yet clear. Here we suggest that even cell division alone is sufficient to produce stresses, which drive the collective migration of cells.
\section{Comparing cell division and intrinsic activity}
\begin{figure*}[htdp]
\centering
\includegraphics[trim = 0 0 0 0, clip, width = 1.0\linewidth]{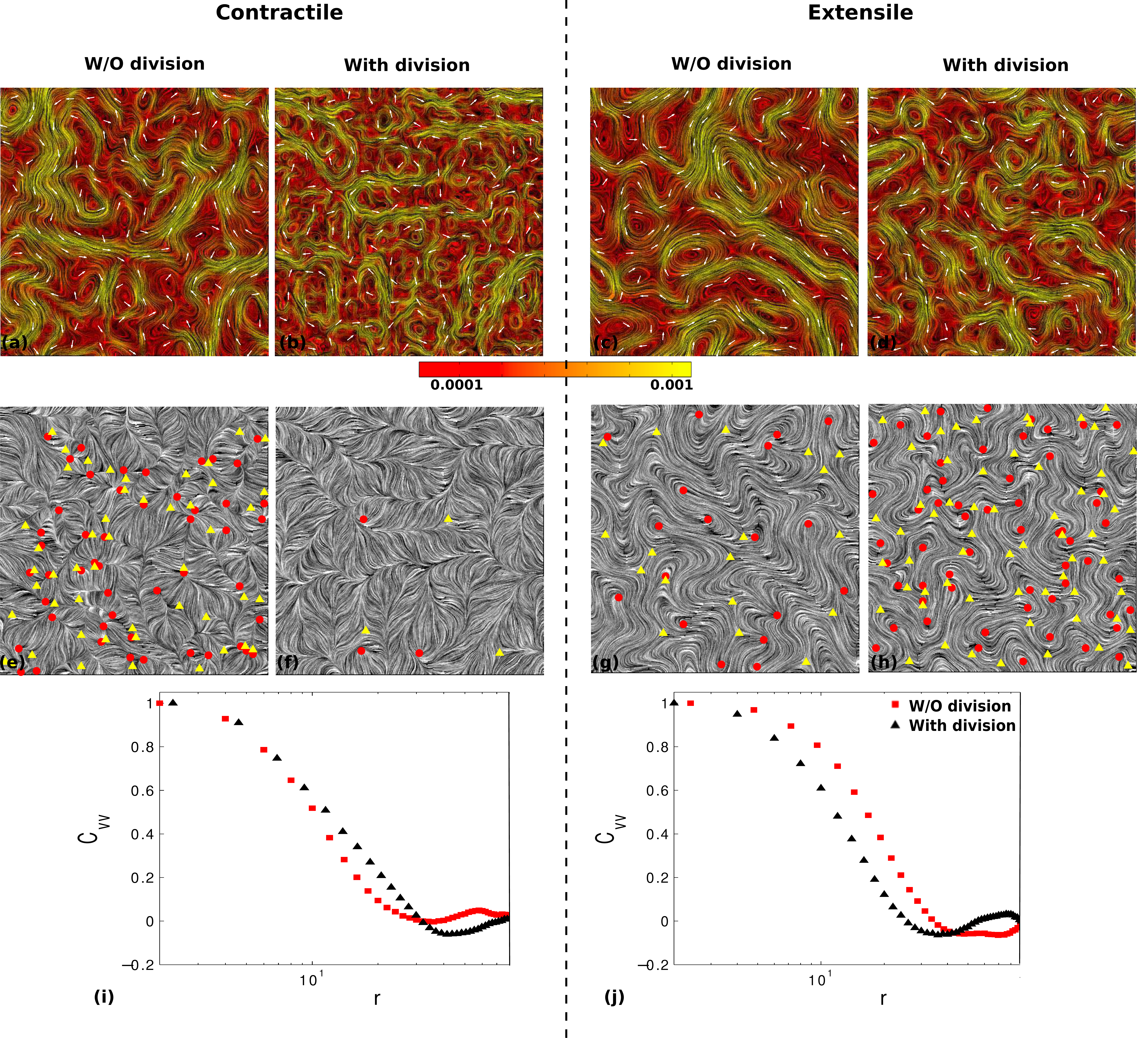}
\caption{The effect of cell division on the dynamics of contractile and extensile cells. (a)-(d), colourmaps show the velocity fields in contractile ((a),(b)) and extensile ((c),(d)) systems (visualisation method is the same as in Fig.~\ref{fig:div}). The colour maps indicate the magnitude of the velocity and white arrows the local direction of the velocity. (e)-(h), colourmaps of the director field indicating  topological defects (with $+1/2$ and $-1/2$ defects denoted by red circles and yellow triangles, respectively). (i),(j), effect of cell division on velocity correlation functions for contractile and extensile cells.}
\label{fig:ex-con}
\end{figure*}
In our discussions so far, the system operates without extensile or contractile stresses due to the intrinsic activity of the cells and the sole source of activity is the cell division. In order to characterise the combined effects of cell division and the intrinsic activity, we next investigate the impact of cell division on the dynamics of an assembly of active cells ($\zeta\ne 0$). The underpinning mechanism of cell activity is not completely clear, but the prevalent microscopic picture is that the activity is generated through the stress that is exerted by myosin motors perturbing the actin cytoskeleton and by the polymerisation of the actin filaments\cite{Myosin1995}. Whether the stress is extensile or contractile for all type of cells is not yet known, but a number of studies show that the myosin contractility is an effective mechanism of propulsion \cite{Rubinstein2009,Cates2012}. Moreover, recent theoretical predictions suggest that the cell activity can show extensile or contractile behaviour depending on the strength of the actin$-$myosin interactions \cite{Voigt2015}. 

A comparison of the flow fields of cells with and without division, shows a notable difference between extensile and contractile cells (Fig.~\ref{fig:ex-con}). Figure~\ref{fig:ex-con}(a) shows the active turbulent behaviour induced by contractile activity of the cells ($\zeta<0$). It is compared with Fig.~\ref{fig:ex-con}(b), where cell division events are incorporated. As discussed earlier, the cell division events are associated with extensile stresses. The extensile and contractile contributions act in opposition to reduce the net stress in the system. The rms velocity falls by $\sim 70\%$ corresponding to a much reduced effective activity. In contrast, introducing division to a culture of extensile cells, the two contributions to the activity add up resulting in an increase in the stress, and the rms velocity inceases by $\sim 125\%$ (Fig.~\ref{fig:ex-con}(c),(d)).

The cell division locally generates vortical structures. These smaller vortices thus appear inside the larger swirls generated by the effective activity, producing an assembly that is characterised by large jets and swirls, which are interleaved by small vortices generated by cell division. This is a characteristically different flow pattern than that observed in meso-scale turbulence due purely to the intrinsic activity \cite{ourepl2014}.

We also calculate the correlation functions for velocity fields (Fig.~\ref{fig:ex-con}(i),(j)). The effect of cell division on the velocity correlation length shows opposite trends in contractile and extensile assemblies: the correlation length is increased in the former (Fig.~\ref{fig:ex-con}(i)), while it is reduced in the latter (Fig.~\ref{fig:ex-con}(j)). This follows the same trend as recent results for extensile active nematics, driven by the intrinsic activity, in the absence of an ordering free energy which found that the correlation length increases (decreases) with decreasing (increasing) activity \cite{ourprl2015}. However, here the variation of the cell concentration due to division events introduces new dynamics to the system and any analogy must be treated with caution.

Another important consequence of cell division is in changing the number density of topological defects, which may be important in controlling the structure of cell layers. Recent experiments on fibroblasts cells have shown that the nematic order of the cells is accompanied by formation of topological defects, which prevent the development of infinite size nematic domains \cite{Silberzan2014}. In the simulations, the number density of topological defects increases with increase in cell division in a system with no intrinsic activity ($\zeta=0$). However, the cell division has a different effect when it is associated with systems having intrinsic extensile or contractile activities. While the generation of topological defects is enhanced by cell division in extensile systems (Fig.~\ref{fig:ex-con}(g),(h)), it is significantly reduced in contractile assemblies (Fig.~\ref{fig:ex-con}(e),(f)). This again can be explained by the division reducing (increasing) the effective activity of contractile (extensile) active nematics in accord with recent studies showing that the number of defects increases with increase in the activity of the system \cite{ourpta2014}. Taking the effects of cell division, a decrease (increase) in the number of defects for contractile (extensile) systems is observed since effective activity is reduced (increased). 
Less topological defects implies less stress in the tissue, which might have important physiological implications.
\section{Cell division and the free surface}
\begin{figure}[tdp]
\centering
{\includegraphics[trim = 0 0 0 0, clip, width = 1.0\linewidth]{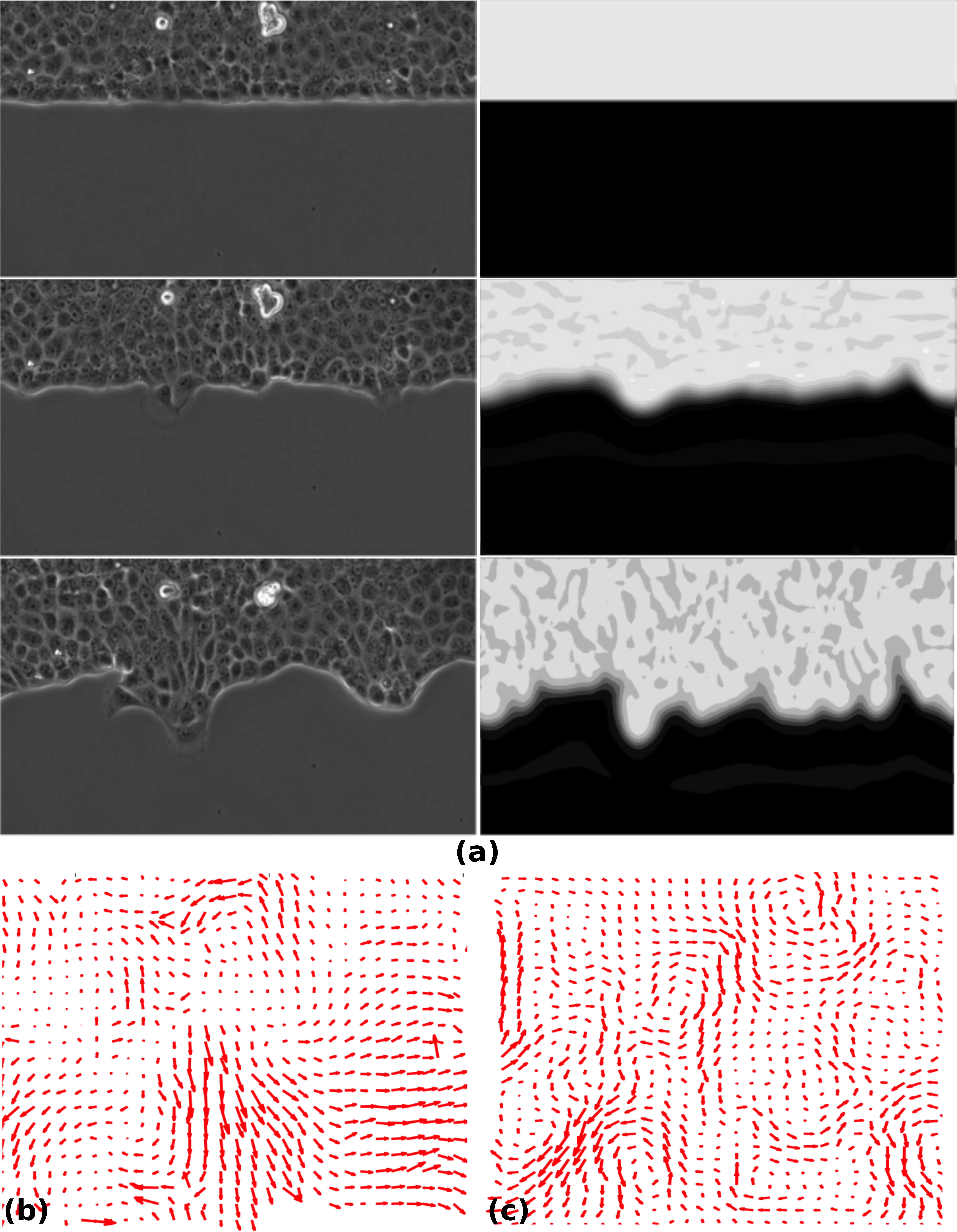}}
\caption{(a) Temporal evolution of a free surface of MDCK cells and emergence of the fingering instability in experiments (left) and the same phenomena observed in our simulations by division-induced activity (right). In the simulations, colormaps show the concentration. The time step in experiments is $150~min$ and in the simulation it is $120$ in simulation units. (b), (c). A close-up of the velocity field in the band for experiment and simulation, respectively.}
\label{fig:band}
\end{figure}
Up to now, we have considered division effects on the dynamics of cell assemblies in periodic domains. In many physiological applications such as morphogenesis, tissue expansion, and wound healing, the mechanical response of a free surface to the cell invasion is of considerable importance \cite{Ghosh2007,Silberzan2007,Wolgemuth2011,Tse2012}. Here, using the equations of lyotropic active nematics, we extend our results to the case where a cell assembly is separated from an otherwise isotropic liquid by a free interface. To distinguish the cell culture from the isotropic fluid, we define a scalar order parameter $\phi$, which measures the relative density of each component with $\phi=1$ for the cells and $\phi=0$ for the isotropic fluid and evolves according to the Cahn-Hilliard equation~\cite{Cahn1958}
\begin{align}
\partial_{t}\phi+\partial_{i}(u_{i}\phi)=\Gamma_{\phi}\nabla^{2}\mu+\alpha\phi,
\label{eqn:conc2}
\end{align}
where $\Gamma_{\phi}$ is the mobility, $\mu=\delta \mathcal{F}/\delta\phi$ is the chemical potential and the free energy of the system is
\begin{align}
\mathcal{F} &=  \frac{A_{\phi}}{2}\phi^{2}(1-\phi)^{2}+\frac{1}{2}A(q^{2}\phi-\frac{1}{2}Q_{ij}Q_{ji})^{2}\nonumber\\
&+\frac{1}{2}\kappa_{\phi}(\partial_{k}\phi)^{2}+\frac{1}{2}K(\partial_{k}Q_{ij})^{2},
\end{align}
where $A_{\phi}$ and $\kappa_{\phi}$ are material constants. Equation~(\ref{eqn:conc2}) together with equations~(\ref{eqn:lc})--(\ref{eqn:ns}) are solved here to describe the dynamics of a dividing colony of cells with free surfaces.
An additional term $\Pi_{ij}=(\mathcal{F}-\mu\phi)\delta_{ij}-\partial_{i}\phi(\partial\mathcal{F}/\partial(\partial_{j}\phi))$ must be added to the stress components in eqn.~(\ref{eqn:ns}), when the variable $\phi$ is introduced. More details of the form of the free energy and the governing equations of lyotropic active nematics can be found in \cite{MatthewPRL}. We use $\Gamma_{\phi}=0.1$, $A_{\phi}=0.08$ and $\kappa_{\phi}=0.01$.
We do not explicitely include any terms in the free energy that lead to interface anchoring \cite{Das2004}, but active anchoring may result from hydrodynamic stresses at the interface \cite{MatthewPRL}.\\
\indent In Fig.~\ref{fig:band}(a), numerical results for the time evolution of the surface of a cellular layer are compared to the results of experiments on the growth of the surface in a colony of dividing MDCK cells. Unlike the experiments, we consider cells with no intrinsic activity ($\zeta=0$) in the simulation to show that a similar behaviour follows from considering the division-induced activity alone. Previous studies have predicted that existence of source terms such as material production can drive hydrodynamic instabilities in the form of undulations at the interface between a viscous fluid and viscoelastic material \cite{Basan2011}. As evident from Fig.~\ref{fig:band}(a), the expansion of the band is accompanied by instabilities that lead to the formation of fingers at the surface in both experiment and simulation. Although previous studies have associated the fingering instabilities to the formation of leader cells at the border \cite{Silberzan2010}, our results suggest that the same phenomena can be induced due to the instability of the nematic field to division-induced activity. It is well known that the presence of activity can result in the formation of bend instabilities in extensile active nematics \cite{Sriram2010,MatthewPRL}. Since the cell division introduces extensile stresses to the cell culture, it can lead to the instability of the nematic field of the cells and induce instabilities at the surface. In addition to the emergence of fingering instabilities at the surface, long-range velocity fields are generated within the growing band even far away from the surface (Fig.~\ref{fig:band}(b),(c)). The appearance of long-range velocity fields with no preferred orientation towards the free surface has been reported in previous studies of tissue growth in response to a model wound \cite{Silberzan2007}. However, in explaining the experimental observations, the emergence of long-range velocity fields and their correlations with the cell movements were attributed to `cryptic'  lamelliopodia, which spread underneath other cells during the tissue growth \cite{Fenteany2005,Silberzan2007}, while here the collective motion is induced by cell division.
Taken together, the comparison of our simulations with experimental observations on the dynamic evolution of the free surface of a cell culture shows that similar qualitative behaviour such as fingering instabilities at the border and long-range velocity fields can be induced by the cell division-induced activity.
\section{Conclusions}
To conclude, we propose a modeling framework that describes the effect of cell division on the dynamics of cell cultures and demonstrate that the model reproduces the experimentally measured flow field around a dividing cell in a MDCK cell culture. We show that an extensile active stress can naturally arise from the nematohydrodynamic representation of cells due to a local increase in cell concentration. We demonstrate that even in the absence of active forcing due to intrinsic activity of the cells, the cell division alone can lead to a coordinated motion of cells. The results suggest that cell division can be considered as one regulator of activity in cultures of extensile and contractile cells. Moreover, we show that the dynamic evolution of a free interface due to the division-induced activity alone (without the intrinsic activity) resembles the experimental observations of the expansion of the cells resulting in fingering instabilities at the interface, as well as previous results on tissue growth and wound healing \cite{Silberzan2007}.

The concept of division-induced activity leads to a broad range of questions about the mechanics of growth in cellular assemblies, for example the escape of cellular layers into the third dimension. The emergence of collective motion due to the active stress generated by cell division alone can stimulate new mechanisms for the control and guidance of cell migration. Our predictions suggest experiments aimed at elucidating the mechanical effects of cell division on salient features of cell assemblies such as the emergence of glassy behaviour at large concentration of cells and the propagation of waves during tissue growth. Future studies will focus on a direct comparison of the simulations and experiments to further characterize the division effect on flow structure and topological defects. For example, a recent study has shown that the frictional damping of the momentum on the scale of hydrodynamic screening length, $\sim\sqrt{\eta/\rho\gamma}$, set by the competition of viscosity $\eta$ and friction coefficient $\gamma$, can play an important role in the dynamics and pattern formation in active nematics \cite{ourNM2015}. The substrate friction will be an important contribution to the dynamics of cellular layers that we hope to investigate in the future. 
\section*{Acknowledgments}
We acknowledge funding from the ERC Advanced Grant MiCE. Financial support from the Mechanobiology Institute is gratefully acknowledged. B.L. acknowledges the Institut Universitaire de France (IUF) for its support. We thank Anh Phuong Le for experimental support, and Arnold Mathijssen, Tyler Shendruk, Matthew Blow, Pascal Silberzan and Lene Oddershede for helpful discussions.




\providecommand{\noopsort}[1]{}\providecommand{\singleletter}[1]{#1}%
\providecommand*{\mcitethebibliography}{\thebibliography}
\csname @ifundefined\endcsname{endmcitethebibliography}
{\let\endmcitethebibliography\endthebibliography}{}

\bibliographystyle{rsc} 

\end{document}